\documentclass{aa}  

\usepackage{graphicx}
\usepackage{txfonts}
\usepackage{multirow}
\usepackage{natbib}
\def\gsim{\;\lower.6ex\hbox{$\sim$}\kern-6.7pt\raise.4ex\hbox{$>$}\;}
\def\lsim{\;\lower.6ex\hbox{$\sim$}\kern-6.7pt\raise.4ex\hbox{$<$}\;}

\def\bmi{\hbox{\it B--I\/}}

\def\Min{${}^{\prime}$\llap{.}}
\def\Sec{${}^{\prime\prime}$\llap{.}}

\def\sec{${}^{\prime\prime}$}

\begin{document}

   \title{Weak Galactic Halo--Fornax dSph Connection from RR Lyrae Stars}

   \subtitle{}

   \author{G. Fiorentino\inst{1}
          \and
          M. Monelli\inst{2,3}
          \and
          P.B. Stetson\inst{4}
          \and
          G. Bono\inst{5,6}
          \and
          C. Gallart\inst{2,3}
          \and
          C. E. Mart{\'i}nez-V{\'a}zquez\inst{2,3}
          \and
          E. J. Bernard\inst{7}
          \and
          D. Massari\inst{1,8}
          \and
          V. F. Braga\inst{5}
          \and
          M. Dall'Ora\inst{9}
          }

   \institute{INAF-Osservatorio Astronomico di Bologna, via Ranzani 1, 40127, Bologna, Italy\\
              \email{giuliana.fiorentino@oabo.inaf.it}
         \and
            IAC-Instituto de Astrof{\'i}sica de Canarias, E-38205 La
            Laguna, Tenerife, Spain
         \and
            ULL-Universidad de La Laguna, Dpto. Astrof\'isica, E-38206 La Laguna, Tenerife, Spain
         \and   
            Herzberg Astronomy and Astrophysics, National Research Council Canada, 5071 West Saanich Road, Victoria, BC V9E 2E7, Canada
         \and
            Department of Physics and Astronomy, University of Victoria, 3800 Finnerty Road, Victoria, BC V8P 5C2, Canada
         \and   
            Dipartimento di Fisica, Universit\`{a} di Roma Tor Vergata, Via della Ricerca Scientifca 1, 00133 Roma, Italy
         \and
            INAF-Osservatorio Astronomico di Roma, Via Frascati 33, 00040 Monteporzio Catone, Italy
         \and
            Laboratoire Lagrange, Observatoire de la Cote d'Azur, F-06304 Nice Cedex 4, France
         \and
            University of Groningen, Kapteyn Astronomical Institute, NL-9747 AD Groningen, Netherlands
         \and
            INAF-Osservatorio Astronomico di Capodimonte, salita Moiariello 16, 80131, Napoli, Italy
             }

   \date{Received August 08, 2016; accepted Y}

% \abstract{}{}{}{}{} 
% 5 {} token are mandatory
 
  \abstract
   {}
   {For the first time accurate pulsation properties of the ancient
     variable stars of the Fornax dwarf spheroidal galaxy (dSph) are discussed in the broad context of galaxy formation and evolution.}
   {Homogeneous multi-band $BVI$ optical photometry of spanning {\it twenty} years has allowed us to identify and characterize more than 1400 RR Lyrae stars (RRLs) in this galaxy.}
   {Roughly 70\% are new discoveries. We investigate the period-amplitude distribution and find that
Fornax shows a lack of High Amplitude
(A$_V\gsim$0.75 mag) Short Period fundamental-mode RRLs (P$\lsim$0.48
d, HASPs). These objects occur in stellar populations more metal-rich
than [Fe/H]$\sim$--1.5 and they are common in the Galactic halo (Halo)
and in globulars. This evidence suggests that old (age older than 10 Gyr) Fornax stars are relatively
metal-poor. 

A detailed statistical analysis of the role of
the present-day Fornax dSph in reproducing the Halo period distribution
shows that it can account for only a few to 20\% of the Halo when  
combined with RRLs in massive dwarf galaxies (Sagittarius dSph, Large 
Magellanic Cloud). This finding indicates that Fornax-like systems 
played a minor role in building up the Halo when compared with massive 
dwarfs.}
   {We also discuss the occurrence of HASPs in connection with the luminosity and the early
chemical composition of nearby dwarf galaxies. We find that,
independently of their individual star formation histories, bright 
(M$_V\lsim$-13.5 mag) galaxies have HASPs, whereas faint ones 
(M$_V\gsim$-11 mag) do not. Interestingly enough, Fornax belongs to 
a luminosity range (--11$<$M$_V<$--13.5 mag) in which the occurrence of 
HASPs appears to be correlated with the early star formation and chemical enrichment of
the host galaxy.}
   
 \par 


   \maketitle
%
%________________________________________________________________

\section{Introduction}\label{sec:intro}

Our Local Group (LG) includes a sizable sample of dwarf galaxies 
located within the immediate surroundings of the two large spirals, 
the Milky Way (MW) and Andromeda (M31). Modern cosmological simulations 
predict that large galaxies assembled by means of merging processes starting from small
protogalactic fragments \citet{searle78}, \citet{springel06}. 
Dwarf galaxies that we see today may be the
relics of such fragments not cannibalized by the MW potential. Galactic
archeology has been adopted to test 
this theoretical framework, with inconclusive results to date. 
 In particular, the iron and the $\alpha$-element
  distributions \citep[see][for reviews]{venn04,tolstoy09,feltzing13} 
  have been used during the last decade as probes of the possible origin of 
  halo stars. The general conclusion is that there are substantial differences 
  in the chemical signatures of the dwarf galaxies compared to the Halo. 
  Moreover, the chemical enrichment of Halo stars is quite complex and
  indeed $\alpha$-element abundances display a well defined
  dichotomy \citep{schuster12} and a steady decrease when moving from the inner to the outer Halo
  \citep[$R_G\ge$ 40 kpc,][]{fernandezalvar15}. 
  However, recent high-resolution spectra clearly indicate a
  broad similarity between the metal-poor, and hence presumably the
  old, stars in the Halo and in several dwarf spheroidals
  \citep[dSphs,][]{frebel10a,frebel10b,starkenburg13a,fabrizio15,frebel15}. 

In the interpretation of these chemical patterns, one must also 
  take into account the complexity of dwarf galaxy chemical evolution, as clearly 
  disclosed by recent high-resolution spectroscopic investigations on 
  Fornax \citep{lemasle14,hendricks14b}, which confirm and expand on previous medium-resolution
  studies \citep{battaglia08,kirby11,letarte06}. Indeed, the Fornax iron
  distribution peaks  at [Fe/H]=--0.8, but the tails extend into the very 
  metal-poor ([Fe/H]$\sim$--3) and the very metal-rich ([Fe/H]$\sim$0) regime.
  The same conclusion applies to the $\alpha$-elements, for which there is observed 
  a clear enhancement in the metal-poor stars that gradually moves to a well defined 
  depletion in the metal-rich ones. The $\alpha$-element distribution
  also shows a well defined ``knee'' at [Fe/H] $\simeq -1.9$ that suggets an early onset of SNe Ia. This extreme chemical 
  complexity goes hand in hand with a complex star formation history in Fornax 
  \citep{deboer12a,delpino13}. Other dwarf
  satellites are more simple in their range of ages and metallicities \citep{martinezvazquez16a,martinezvazquez16b}, but the comparison between 
  Halo stars and red giant branch (RGB) stars in nearby dwarf galaxies is still hampered by the fact 
  that Halo stellar populations are dominated by old stars, while 
  those in dwarf galaxies suffer the age-metallicity degeneracy and are a mix of 
  old and intermediate-age stars. 

That stars of different ages show a clear 
  dichotomy in their chemistry and kinematics is illustrated by the results on 
  Carina by \citet{fabrizio16}. Based on solid tracers of the old Horizontal Branch (HB) and intermediate-age 
red clump (RC) populations in Carina, these authors find a systematic difference in the metallicity 
distribution and kinematics between these two
sub-populations. Specifically, current metallicity estimates based on
RGB stars mainly trace the intermediate-age (RC) 
population, whereas the old stars are systematically more metal-poor.  
These considerations are consistent with the evidence that the intermediate-age star formation
  episode in Carina accounts for more than 70\% of its stellar
  content \citep{dolphin02a,small13}.

%~~~~~~~~~~~~~~~~~~~~~~~~~~~~~ FIG 1 ~~~~~~~~~~~~~~~~~~~~~~~~~~~~~~~~~~~
\begin{figure}
\centering
%\vspace{-5cm}
	\includegraphics[width=8cm]{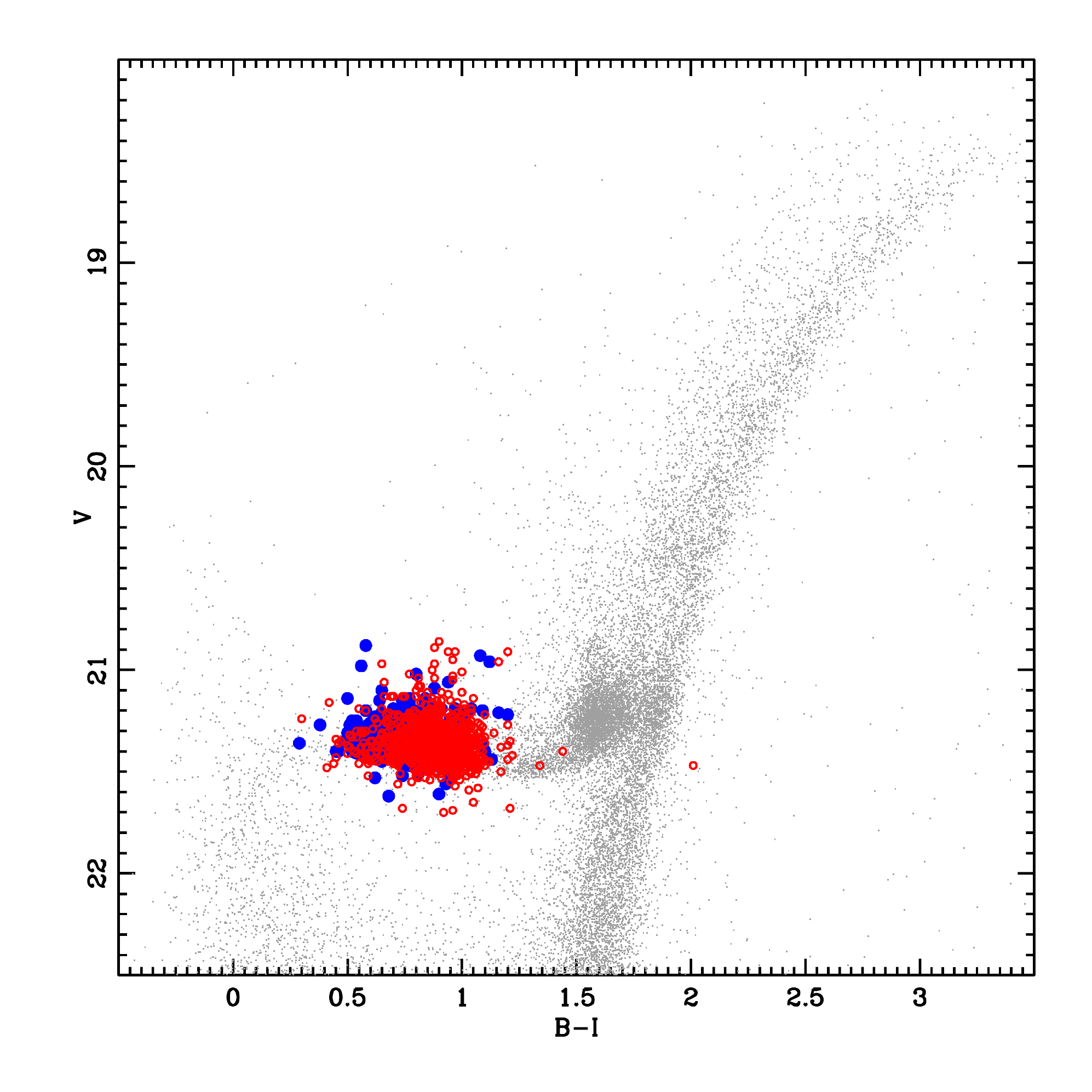}
%#\vspace{-1cm}
    \caption{The V, B--I colour-magnitude diagram. Blue dots and red
      open circles mark RR$c$ and RR$ab$, respectively.}
    \label{fig1}
\end{figure}
%~~~~~~~~~~~~~~~~~~~~~~~~~~~~~~~~~~~~~~~~~~~~~~~~~~~~~~~~~~~~~~~~~~~~~~~

An ideal experiment is one that compares stellar populations with the same age in
both the Halo and nearby dwarf galaxies. This opportunity is offered by 
RR Lyrae stars (RRLs). They are precise distance indicators  
and solid tracers of evolving stars with initial mass lower than $\sim\,$0.8 M$_\odot$, i.e., 
stars older than 10 Gyr. Even though in extragalactic systems these stars are too faint
($\sim\,$3.5-4~mag fainter than the tip of the RGB) to spectroscopically measure their 
individual chemical compositions, they can easily be detected and 
characterized by their pulsation properties (mean magnitudes, 
amplitudes, periods) well beyond our LG \citep[e.g., the RRLs observed 
at $\sim$ 2 Mpc in the Sculptor group][]{dacosta10}. 
They typically pulsate in the fundamental mode (RR$ab$), the first overtone
(RR$c$), or both.\par 
In recent years, different photometric sky surveys have provided the 
pulsation properties of a large number ($\sim$ 15'000) of Halo RRLs 
\citep[e.g., CATALINA survey][]{drake13}. Moreover they are ubiquitous, and indeed 
they have also been detected in galaxies of modest mass: 
classical dwarfs \citep[e.g.][]{coppola15,martinezvazquez16b} and ultra-faint 
dwarfs \citep[e.g.][]{garofalo13,boettcher13}. Our group is systematically 
investigating variable stars in the LG  galaxies using both proprietary and archival data
in order to maximize temporal coverage. The global photometry, including the properties of
the variable stars are available to the community (e.g., Leo~I, Carina, Sculptor\footnote{\url{http://www.cadc-ccda.hia-iha.nrc-cnrc.gc.ca/en/community/STETSON/homogeneous/}}).\par
In this investigation we present new results for the RRLs in the Fornax dSph. These are used to provide an independent 
and global picture of the early chemical enrichment and the star-formation history 
of this interesting galaxy. Fornax is among the most massive galaxies surrounding the MW
and one of the few gas-poor stellar systems of the LG hosting
(five) GCs. Although Fornax is relatively close 
\citep[$\mu_0=20.84\pm$0.04 mag,][using the tip of the RGB]{pietrzynski09} and it has been the target of extensive
photometric \citep{coleman08,deboer12b} and spectroscopic
\citep{letarte10,kirby11} investigations, we still lack a complete census of evolved variables.
The main previous investigation was performed by \citet{bersier02}
and covered an area of nearly half a square degree, a minor fraction of the body of
the galaxy \citep[r$_t=$71 arcmin][]{mateo98a}. Their {\it V, I\/} time-series data were collected using 1-1.3m 
telescopes in $\sim\,$2\sec seeing, and provided a sample of 396 RR$ab$
and 119 RR$c$ as classified on the basis of their periods. The quality
of their photometry was inadequate to determine accurate light curves, and the amplitudes of 
variation were not provided. A further, but still preliminary, investigation
of cluster and field variables in Fornax was provided by
\citet{greco05} using the imaging subtraction method. They found
more than 700 candidates (see their Table 2) among RR Lyrae, Anomalous
Cepheids and Dwarf Cepheids (SX Phoenicis). However, a detailed
comparison is hampered by the lack of finding charts and astrometric
positions of the candidate variables.

In this paper we present a nearly complete sample of RRLs observed in
Fornax dSph (Section~\ref{sec:phot}). For the first time we show their
amplitude vs period diagram and we discuss
its implications for the early chemical enrichment of Fornax dSph (Section~\ref{sec:bailey}). The period distribution of fundamental mode
RRLs in Fornax dSph is used to derive constraints on the contribution
of Fornax-like systems to the Galactic halo formation when compared with more massive dwarfs
(Sagittarius and Large Magellanic cloud systems, hereinafter Sgr and
LMC respectively), see Section~\ref{sec:discussion1}. Finally, we interpret these findings in the
broad context of LG galaxies (Section~\ref{sec:discussion2}).

%_____________________________________________________________________________________________

%~~~~~~~~~~~~~~~~~~~~~~~~~~~~~ FIG 2 ~~~~~~~~~~~~~~~~~~~~~~~~~~~~~~~~~~~
\begin{figure*}
\centering
	\includegraphics[width=8cm]{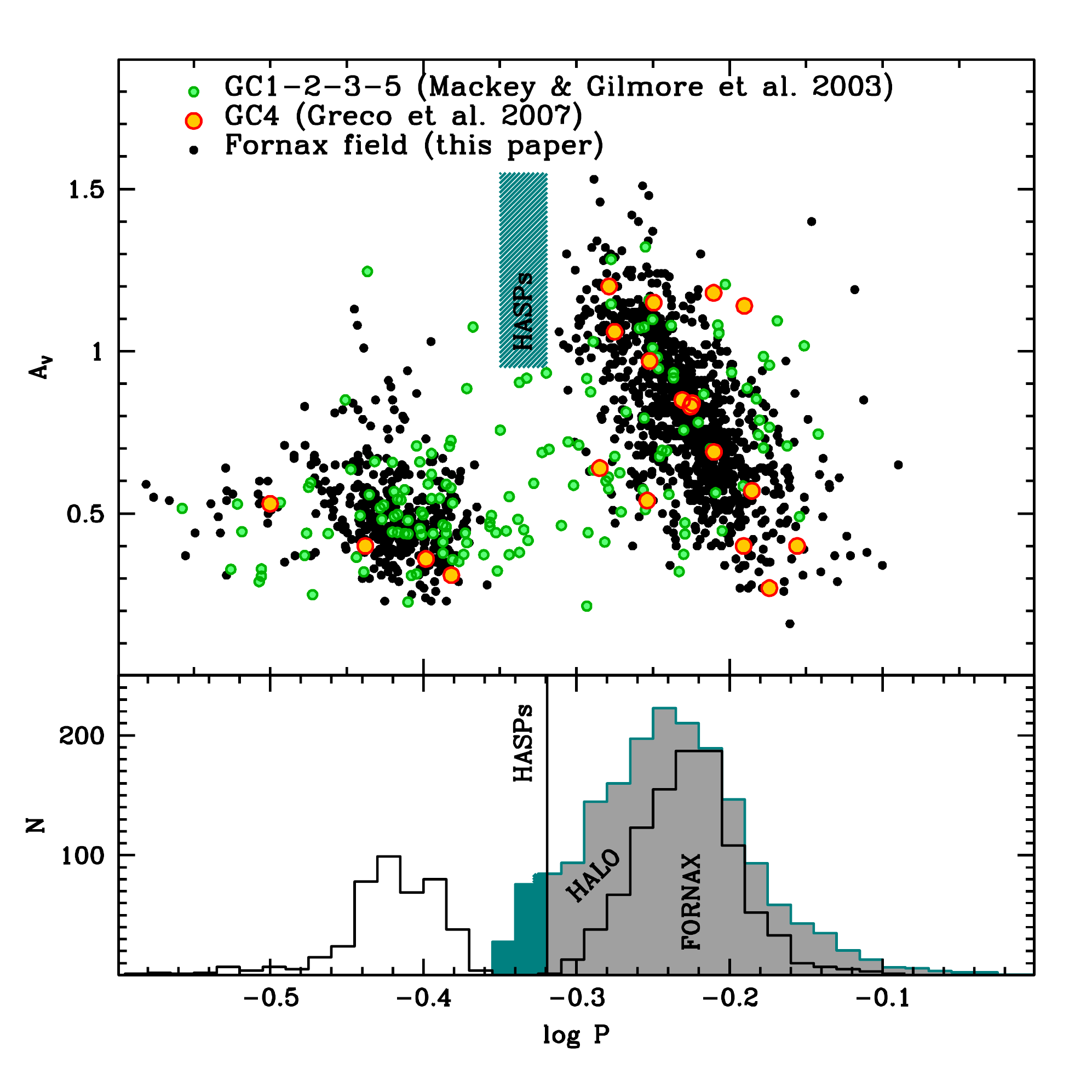}
	\includegraphics[width=8cm]{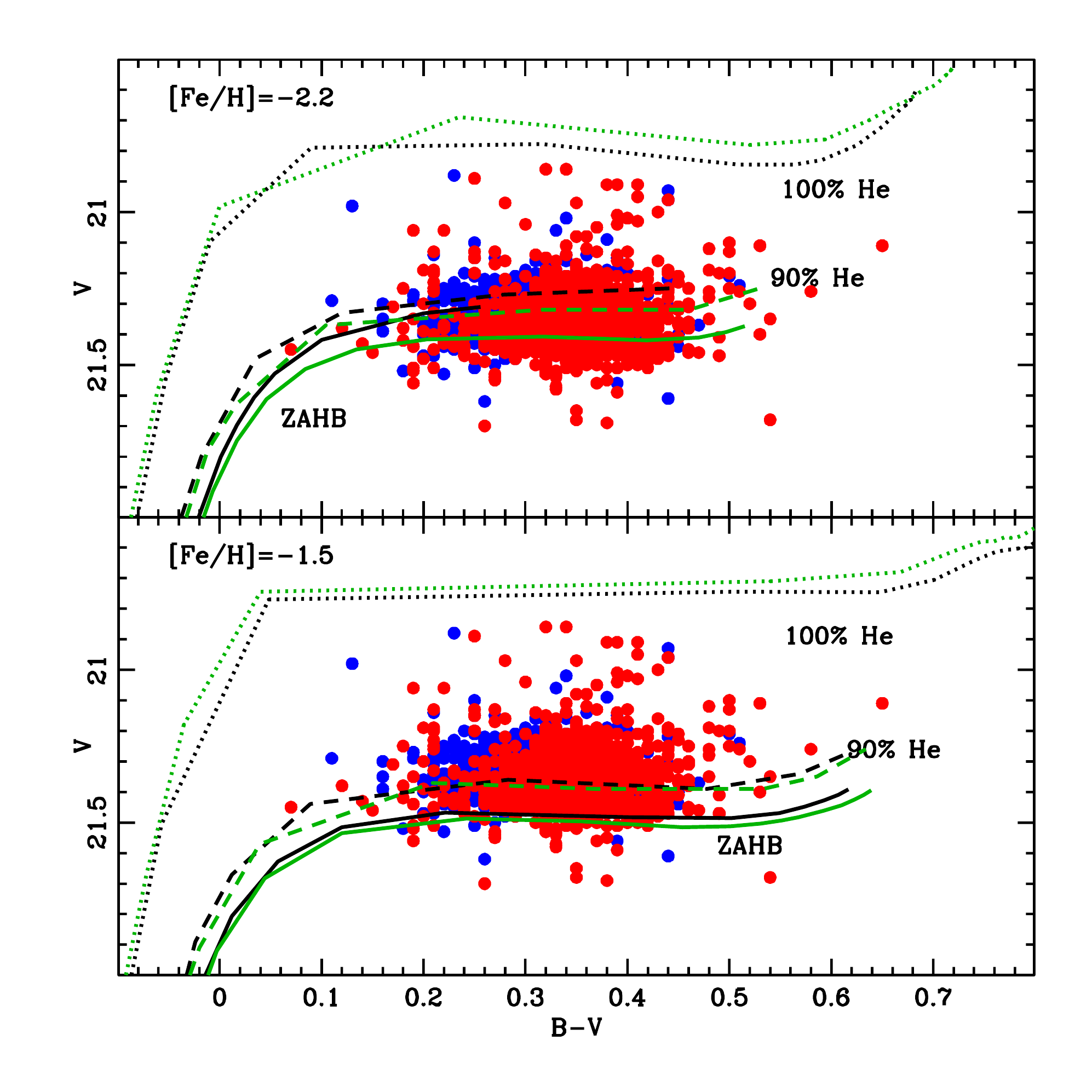}
    \caption{{\it Left panels:} The period-V band amplitude (Bailey) diagram (top)
      and the period distribution (bottom) of the 1426 Fornax RRLs
      (black dots). RRLs in the Fornax globular clusters are also
      shown. The teal coloured region indicates the HASPs'
      location. The grey/teal coloured histogram is the period distribution of $\sim$15'000 Halo RRLs. {\it Right panels:} Zoom on the Horizontal Branch of the Fornax dSph is shown as compared with theoretical
      predictions for scaled-solar (black lines) and alpha-enhanced
      (dark green lines) chemical compositions. The loci of
      the Zero Age Horizontal Branch (solid lines), of the stars that
      have burned 90\% (dashed) and 100\% (dotted) of their central Helium are shown for two selected metal abundances.
}
    \label{fig2}
\end{figure*}
%~~~~~~~~~~~~~~~~~~~~~~~~~~~~~~~~~~~~~~~~~~~~~~~~~~~~~~~~~~~~~~~~~~~~~~~

\section{Photometric Data sets and RR Lyrae identification}\label{sec:phot}

The multi-band ($BVI$) optical photometry employed here is 
based on both proprietary and archival datasets.  
The images were collected with telescopes ranging from 0.9m to 8m aperture
and cover a time interval of nearly 20 years (1994 December to 2014 October). The median seeing 
was 1\Sec 1, with an interquartile range of 0\Sec 95-1\Sec 3.
The data reduction and the calibration is part of the effort of PBS to build 
a data base of homogeneous photometry including star clusters
and resolved galaxies. A full description of the data, data reduction, and of all
the variable stars identified will be addressed in a future paper.

The variable candidates for which we have been able to determine at least a provisional period cover a sky area of 
$\sim\,$50\Min$\times$54\Min\ around the galaxy center. The number of measurements 
per band is on average about 130 in $B$, 180 in $V$ and 30 in $I$. 
Our approach for identifying variable stars has been 
described in several previous investigations 
\citep[][]{stetson14a,stetson14b}. 
We have identified $\sim$ 1900 candidate variables, among
them 1426 bona fide RRLs for which we can fit light curves. In this investigation we
focus on this unique sample of RRLs, 
since it increases the known sample by a factor of nearly
three. Fig.~\ref{fig1} shows the location of the current RRL
sample in the $V$, {\it B--I\/} Colour-Magnitude diagram. 
The accuracy and the precision of this photometry is 
highlighted by the smooth transitions among the different populations
in the region of the HB (V$\sim$21-22). One can observe the blue
HB ($V\,\sim\,$22.0-21.5, \bmi$\sim$0-0.5 mag), the RRLs
($V\,\sim\,$21.0-21.5, \bmi$\,\sim\,$0.5-1.2~mag), the red HB ($V\,\sim\,$21.0-21.5, 
\bmi$\,\sim\,$1.2-1.5~mag) representing the old (age older than 10 Gyr) stellar
population and the RC stars ($V\,\sim\,$20.9-21.4, \bmi$\,\sim\,$1.5-1.8~mag) 
tracing the intermediate-age population. Furthermore, the RGB stars display a well defined stellar over-density associated with the 
RGB bump ($V\,\sim\,$20.9-21.5, \bmi$\,\sim\,$1.7-1.9).  
Note that we have plotted flux-weighted mean magnitudes for the variable strs.
The few outliers observed can be explained by imperfect
sampling of the light curve in at least one of the three bands
(generally the $I$-band). This
also explains why there are several RR$ab$ (red open circles)
in the bluer region of the instability strip, 
i.e., the region where typically only RR$c$ variables (blue dots) 
are located.

\section{The Bailey diagram}\label{sec:bailey}

To provide a solid classification of the RRLs in our sample we adopted 
the Bailey diagram (luminosity amplitude versus period), shown in 
Fig.~\ref{fig2} (top left panel). Data (black dots) plotted in this figure display a well defined transition between RR$c$ and 
RR$ab$ in the period range between $\log P$=--0.35 and $\log P$=--0.32 d. 
On the basis of this criterion we end up with 436 RR$c$ and 990 RR$ab$ variables.  
It is worth noting that Fornax lacks High Amplitude 
Short Period RR$ab$ variables (HASPs), i.e., RR$ab$ with periods shorter than 
P$\lsim$0.48 d\footnote{The boundary of the HASP at P$\lsim$0.48$\,$d
  \citep{fiorentino15a} was arbitrary and based on preliminary
  evidence that no small dSph of the MW was known to host RRLs of this kind. Importantly, the updated catalogues for
Sculptor \citep{martinezvazquez15} and Fornax (this work) confirm that choice.
The precise value of this limit has no impact on the present
discussion.} \citep[teal coloured region in Fig.~\ref{fig2}, left panel, see also][]{fiorentino15a}. Note that the lack of HASPs is also evident 
among the RRLs hosted in the five Fornax GCs \citep[green and red
circles][respectively]{mackey03,greco07}.

In this context it is
  worth mentioning that across the transition region between RRc and
  RRab variables are also found the mixed-mode (RRd) variables that pulsate simultaneously in the first overtone
  (typically the main mode) and in the fundamental mode. They display
  a well-defined amplitude modulation but this is typically of the
  order of a few tenths of a magnitude \citep{smolec16}. In this
  region are also found Blazhko RRLs, i.e., RRLs that display
  amplitude and/or phase modulations on a time scale ranging from tens
  of days to years. The amplitude modulation ranges from a few tenths
  to more than half a magnitude \citep{piersimoni02,braga16}. The
  pulsation characteristics available in the literature do not allow
  us to assess whether the Fornax cluster RRLs located below the HASP region
  (green circles in Fig.~\ref{fig2}, left panel) are either RRd or
  Blazhko RRLs. Therefore, the periods and amplitudes for these stars, that appear
to have in principle the right periods to be classified as HASPs, may
not be definitive since the data presented in
\citet{mackey03} are undersampled, as discussed in their Section~3.4. 
However, four out of the five Fornax GCs 
are metal-poor and thus not expected to produce HASP RRLs (see below). These are the 
reasons why we do not further discuss
the GC sample.

Circumstantial empirical and theoretical evidence indicates that the lack 
of HASPs can be explained by a lack of old stellar populations more 
metal-rich than [Fe/H]$\sim$--1.5 \citep[as shown in][]{fiorentino15a}. 
This finding may appear to be at odds with current spectroscopic measurements 
\citep{battaglia08,kirby11,letarte06,letarte10,lemasle14,hendricks14b} 
showing that Fornax RGB stars peak at [Fe/H]$\sim$--0.9 and range from 
--3.0 to --0.4 dex.  Clearly, more metal-rich populations are present in 
Fornax, but they may just be too young to produce RRLs. A number 
of studies \citep[e.g.]{fabrizio15,fabrizio16,martinezvazquez16b} 
have demonstrated that purely old stellar populations do not necessarily 
share the mean chemical abundance of RGB stars.  
In agreement, the globular clusters support the inference that the old 
populations in Fornax are significantly more metal-poor than the average: 
the mean metallicity of four out of the five clusters is [Fe/H]=--2 or
less \citep{letarte06}, and only GC 4 has a mean metallicity as high
as of [Fe/H]$\sim$--1.40$\pm$0.06
(\citealt{larsen12a} using integrated-light,
high-dispersion spectra, though it is also believed
to be 3 Gyr younger than the other four, \citealt{buonanno99}).
We note here that \citet{letarte06} found a bias in the estimated metallicity to
higher values (amounting to $\sim$+0.3 to +0.5 dex) when using integrated-light spectra instead of
individual stellar measurements.
%~~~~~~~~~~~~~~~~~~~~~~~~~~~~~ FIG 3 ~~~~~~~~~~~~~~~~~~~~~~~~~~~~~~~~~~~
\begin{figure}
\centering
	\includegraphics[width=8.cm]{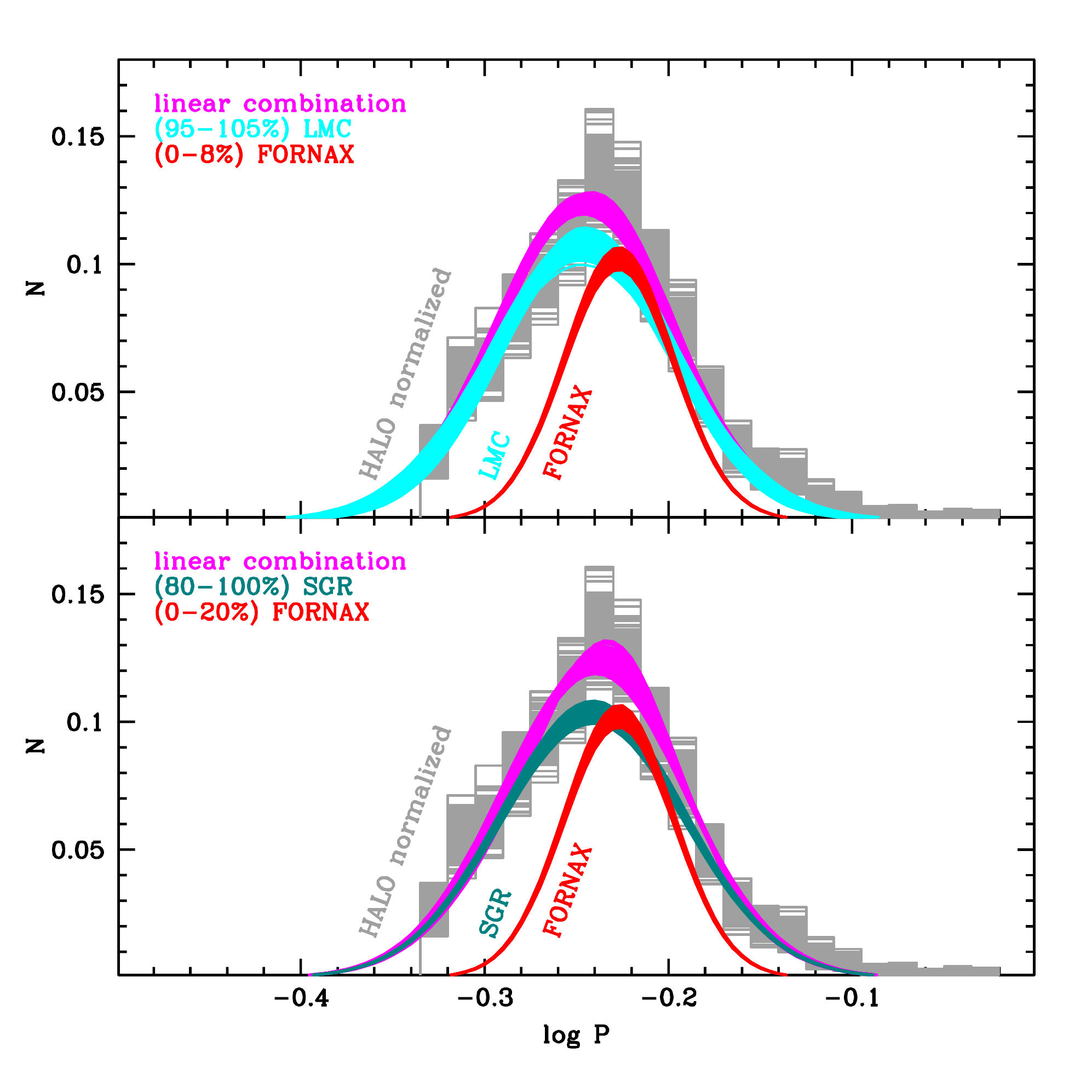}
    \caption{Galactic halo period distributions for the 200
      realizations of randomly extracted sub-samples (grey
      histograms). With magenta lines we have shown the linear
      combinations of Fornax (red) period distribution with 
      stellar systems that do host HASPs, i.e., LMC (cyan), Sgr (teal
      green). The individual contributions to the Galactic halo for each stellar system used
      for the linear combinations are also given in each panel.
    } 
    \label{fig3}
\end{figure}

To further test the hypothesis that Fornax RRLs are metal poor ([Fe/H]$\lsim$--1.5), Fig.~\ref{fig2} (right panels) shows the comparison 
between the current sample of RRLs and evolutionary predictions for helium-burning 
structures available in the BaSTI data base. To compare theory and
observations we adopted a true
distance modulus of $\mu_0$=20.84$\pm$0.04 mag, a reddening of
0.03~mag \citep[][]{pietrzynski09}
and the \citet{cardelli89} reddening law. The bottom panel of this figure shows the comparison between the 
theoretical Zero Age Horizontal Branch (ZAHB, solid black line) for two different
chemical compositions [Fe/H]=--1.5 (bottom panel) and --2.2 (top panel) and 
for both a scaled-solar and an alpha-enhanced chemical mixture \citep{pietrinferni14}.  
The dashed and dotted lines represent the loci of 90\% and 100\% helium 
exhaustion in the core. The time for the last 10\% of helium exhaustion is only a few percent
compared to the 
total lifetime spent close to the ZAHB (98\%). The models suggest 
that RRLs in Fornax can be hardly explained as an old stellar population 
with a narrow and quite metal-rich ([Fe/H]$\gsim$--1.5
dex) metallicity distribution. A metallicity distribution 
ranging from at least --1.5 dex to less than --2.2 dex seems more
compatible with the wide HB distribution of RRLs. We note that the adopted
evolutionary scenario does not affect our conclusions, indeed alpha-enhanced models for [Fe/H]$\sim$--1.5 predict a ZAHB only slightly
fainter ($\Delta$V$\sim$+0.02 mag, darkgreen lines in figure) than scaled-solar ones.

Our conclusion is supported by an independent analysis by
\citet{deboer12b}. They computed a detailed
star formation history (SFH) and an age metallicity relation (AMR) combining deep- and wide-field photometry with medium
resolution spectroscopy. They find that Fornax is dominated
by intermediate age stars ($<$10Gyr) but also hosts an old population ($>$10Gyr). 
Furthermore, their AMR indicates
that the old population has a metallicity [Fe/H]$\lsim$--1.5 dex, thus
explaining the lack of HASPs in spite of the very large RRL sample.

%~~~~~~~~~~~~~~~~~~~~~~~~~~~~~ FIG 4 ~~~~~~~~~~~~~~~~~~~~~~~~~~~~~~~~~~~
\begin{figure*}
\centering
	\includegraphics[width=11.cm]{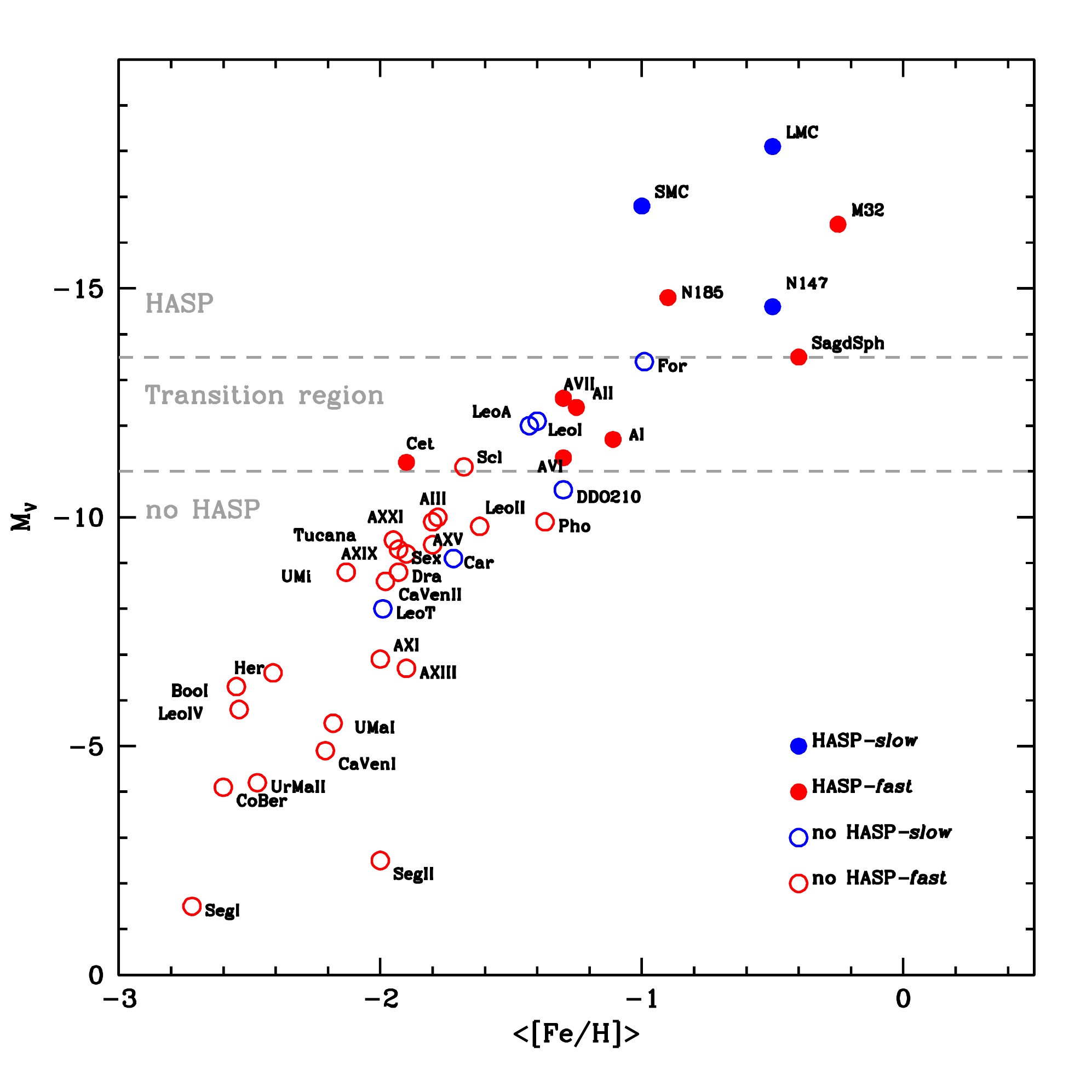}
    \caption{Luminosity-Metallicity distribution \citep{mcconnachie12} for all the dwarf
      galaxies of the LG for which a short-period variability
      study is available. Filled and empty circles mark either the presence or the lack of HASPs. {\it slow} and {\it fast} galaxies are colour coded with red and blue respectively.
    } 
    \label{fig4}
\end{figure*}

%_____________________________________________________________________________________________
\section{Building up the Galactic halo with Fornax-like dSph}\label{sec:discussion1}

In \citet{fiorentino15a} we have shown that HASPs are observed in the
Halo (see teal region in Fig.~\ref{fig2}) while they are typically
absent in dwarf satellites of the MW, with the
exception of the most massive ones, such as Sgr and the
Magellanic clouds. Thus, despite Fornax's large total mass, its huge RRL sample
shows a behavior similar to the majority of low-mass dwarfs. Due to their 
very high amplitudes and to the good sampling of our dataset, the HASP 
deficiency can not be an observational bias and has strong
implications for Halo formation. On the basis of the observational 
evidence that low mass dwarfs do not contain HASP RRL, in
\citet{fiorentino15a} we have argued that a major
 role for small dwarfs, as we see them today, can be excluded in building up the
 Galactic halo. This conclusion was based on a large sample of RRLs for the Halo ($\sim$
  15000 stars) and for small ($\sim$1300 stars) and large (LMC, SMC
  and Sgr) dwarfs. A similar conclusion was achieved by \citet{zinn14} but based on a smaller
sample of Halo field (Smooth Halo, $\sim$ 170 objects) and Fornax
dwarf \citep[$\sim$396 objects,][]{bersier02,greco05}.

In Fig.~\ref{fig2} we show the
period-amplitude diagram and the period distribution of the RRLs in
Fornax (black solid line). For comparison we have plotted the
Halo distribution for RRab stars (grey region)
arbitrarly rescaled. The Fornax period distribution is very different from
that for the Halo and resembles that found using the data on small
mass dwarfs collected by \citet{stetson14b}. The latter is indeed well
described by a Gaussian function peaked around $\sim$0.59 d and with a
small $\sigma$ (0.05). A Kolmogorov-Smirnoff test excludes the possibility that 
the two period distributions (bottom left panel of Fig.~\ref{fig2}) are drawn from the same parent population. 

In order to quantify the role of Fornax-like systems in Halo
formation we have adopted the following approach. The
main difference between the Halo and the Fornax period distribution is seen
in their short-period tail.  Therefore we decided to reproduce the Halo
period distribution using a linear combination of Fornax-like systems
plus massive galaxies that do host HASPs, i.e., the LMC and
Sgr \citep{fiorentino15a}. For the sake of simplicity we have approximated all the
distributions by analytical gaussian functions. To compare samples
with similar size we have randomly extracted $\sim$2000 objects from the full
Halo sample (grey histograms in Fig.~\ref{fig3}). This extraction
was repeated 200 times. We have used the same procedure for the LMC
sample.
For each Halo realization we have computed the linear combination that
best reproduces the Halo gaussian function, see Fig.~\ref{fig3}
(magenta lines). In each panel of this figure we also have shown the
individual components that form the linear combination. Interestingly enough,
when combined with Sgr (teal green lines, bottom-left panel), Fornax can
contribute a fraction that goes from 0 to 20\% of the Halo whereas this
fraction decreases to 0-8\% for the LMC (cyan lines, top-left panel).
This suggests that Fornax-like systems, unlike other massive dwarfs,
likely have mimimally contributed to the Halo. In this sense Fornax
seems to be a transitional system between small and massive dwarfs.

\section{HASPs and the early host galaxy evolution}\label{sec:discussion2}

To further investigate the connection between the chemical enrichment 
history in dwarf galaxies and their HASP content, we extend our
discussion to the LG. Fig.~\ref{fig4} shows the classical mean metallicity-luminosity relation 
\citep[data from][]{mcconnachie12} for LG dwarf galaxies. Only stellar 
systems in which at least one RRL has been
found have been included. Filled circles represent systems hosting at least one HASP, while red and blue
symbols represent {\it fast\/} and {\it slow\/} evolvers in the nomenclature of
\citet{gallart15}: the {\it fast\/} galaxies formed the vast 
majority of their mass in stars at early epochs ($>$10 Gyr), 
while the {\it slow\/} galaxies are mainly dominated by a mix of 
intermediate-age and young stellar populations. This means that 
the old population in the latter systems represents only a
minor fraction of the entire baryonic mass budget.

Data plotted in this figure indicate that the current sample can be split into 
three different groups:\par
{\bf - Bright systems -} this group includes stellar systems brighter than 
$M_V\sim$--13.5 mag, which {\it all} host HASPs. These galaxies surround both the MW (SMC,
\citealt{soszynski10b}; LMC, \citealt{soszynski09a}; Sgr dSph, \citealt{soszynski14}) and M31 
(NGC147, NGC185, Monelli et al. in preparation; M32,
\citealt{fiorentino10a,fiorentino12a}).
With the exception of the Magellanic clouds and
NGC147, the other galaxies are dominated by old stellar populations.\par
{\bf - Faint systems -} this group includes stellar systems fainter than 
$M_V\sim$--11 mag; none of these host HASPs. This is the largest sub-sample,
dominated by gas-poor (dSph, ultra faint) systems made of mostly
old stellar populations. A few galaxies hosting a
significant intermediate-age population are also
present, i.e., Carina \citep{monelli03}; Leo~T dSph/dIrr
\citep{clementini12}; DDO210 dSph/dIrr \citep{cole14}.\par
{\bf - Intermediate luminosity systems -} this group includes stellar systems with 
absolute magnitude ranging from $M_V\sim$--11 to --13.5 mag. These may
or may not host HASPs, and we argue that the presence of HASPs is related to their individual early
SFH (see below).\par 

This figure indicates that, among the more massive stellar systems, 
both {\it slow\/} and {\it fast\/} evolvers
have had an early chemical enrichment that was rapid enough to 
produce such old ($\gsim 10$~Gyr), metal-rich ([Fe/H]$\,\gsim\,$--1.5)
stellar populations that we observe nowadays as HASPs. In contrast,
the less massive galaxies
%, largely dominated by {\it fast\/} -i.e. mainly \ old - systems, 
had an early chemical enrichment insufficient to reach the
necessary metallicity to produce HASPs. In the intermediate
luminosity region there is also a mix of  {\it slow\/} 
and  {\it fast\/} galaxies. However, all the {\it slow\/} evolvers in this
region completely lack HASPs (Leo~I, \citealt{stetson14a}; Leo~A, \citealt{bernard13};
Fornax, this work), while the {\it fast\/} do host a small fraction of them  
($\lsim$ 3\%; Sculptor \citealt{martinezvazquez16b}; Cetus, \citealt[][]{bernard09}; 
Andromeda~I and Andromeda~II, Mart{\'i}nez-V{\'a}zquez et al. in preparation; Andromeda~VI, \citealt{pritzl02}; 
Andromeda~VII, Monelli et al. in preparation). Therefore, {\it in the transition region\/} the total current 
baryonic mass ($\sim$2-20$\times 10^6$, \citealt{mcconnachie12}) 
is not the only parameter driving the early chemical 
enrichment---and, in turn, the presence or lack of HASPs.
The total luminosities of Fornax and Sculptor seem to be
the lower limits for the presence of HASPs among {\it slow\/} and
{\it fast\/} galaxies, respectively. This suggests that the presence/lack of HASPs in
this luminosity range, with HASP present in {\it fast\/} galaxies but 
not in {\it slow\/} ones, depends strongly on the details of the early SFH, and further
strengthens the hypothesis that {\it slow\/} and
{\it fast\/} galaxies were already different $\sim$10 Gyr ago \citep{gallart15}. 
It also establishes a means to get insight on the early SFH of dwarf galaxies based 
on the presence or absence of HASP among their RRL population. \par

Among the four cluster-bearing dwarf galaxies, the Sgr and Fornax
dSphs have the same current luminosity (barionic mass, see Fig.~3) and similar
numbers of globulars \citep{mackey04}. However, their mean metallicities
and their evolution histories are quite different. As a matter of fact, the number of RRab in Sgr 
is almost a factor of two larger than in Fornax ($\sim$1600 {\it vs} 1000) and 
there are clear differences in their RR$ab$ period distributions and in the HASP
occurrence. The role played by Sgr and Fornax in building up the Halo has 
already been discussed by \citet{zinn14}, using more limited samples 
of field RRLs, and by \citet{mackey04} using their cluster RRLs. Our
findings support their results, further stressing the major 
contribution of massive stellar systems like Sgr and/or the LMC when
compared with Fornax-like dSph galaxies ($\sim$80-100\% vs 0-20\%) in
very good agreement with current cosmological simulations \citep{cooper10,tissera14}. In passing,
 we note that this is also consistent with the hypothesis put forward by 
\citet{gallart15} that {\it slow\/} dwarfs such as Fornax may have formed 
in low density environments, away from the main cosmic density fluctuations 
that gave rise to the MW or M31, which implies also a later accretion into the main galaxies.

\section{Final remarks}\label{sec:final}

The use of pulsation properties of RRLs (amplitudes, periods) 
as diagnostics of the metallicity distribution of old stellar 
populations dates back to more than one century ago. The results 
presented in this paper indicate that they can also be adopted to provide
independent constraints for galaxy formation models, and in particular, for the 
role played by mergers in building up the halo of massive
galaxies \citep{bullock05,zolotov09}.

This investigation paves the way to the use of RRLs for tracing the early
star formation and chemical enrichment histories in Local Group galaxies and in
nearby galaxy clusters. This opportunity becomes even more compelling
if we account for the homogeneous multi-band (u,g,r,i,z,y) time series
data that will be collected by the Large Synoptic Survey Telescope 
\cite[LSST][]{ivezic10} and for the deep limiting magnitudes of individual
exposures. Moreover, near-future space (James Web Space Telescope) and ground-based 
Extremely Large Telescopes will offer the opportunity to
extend photometric and spectroscopic investigations of primary
distance indicators to Local Volume galaxies.

\begin{acknowledgements}

This research has been supported by the Spanish Ministry of Economy and 
Competitiveness (MINECO) under the grant AYA2014-56795-P.
EJB acknowledges support from the CNES postdoctoral fellowship program.
GF and DM have been supported by the Futuro in Ricerca 2013 (grant
RBFR13J716).

\end{acknowledgements}

\bibliographystyle{aa} % style aa.bst
%   \bibliography{ms.bib} % your references Yourfile.bib
%\bibliography{/Users/giuliana/work/references/ms}

\end{document}